\documentclass[english,11pt, twocolumn]{iopart}
\usepackage[T1]{fontenc}
\usepackage[latin9]{inputenc}
\usepackage{textcomp}
\usepackage{graphicx}
\usepackage{amssymb}



\usepackage{babel}

\begin{document}

\title{Measurement of small photodestruction rates of cold, charged biomolecules
in an ion trap}

\author{D. Offenberg, Ch. Wellers, C. B. Zhang, B. Roth, and S. Schiller}

\address{Institut für Experimentalphysik, Heinrich-Heine-Universität Düsseldorf,
40225 Düsseldorf, Germany}

\begin{abstract}
In this work, we demonstrate quantitative measurements of photodestruction
rates of translationally cold, charged biomolecules. The long-term
stable storage of the molecular ions in an ion trap at ultra-high
vacuum conditions allows measurement of small rates and verification
that rates are linear in photodestruction laser intensity. Measurements
were performed on singly protonated molecules of the organic compound
glycyrrhetinic acid (C$_{30}$H$_{46}$O$_{4}$), dissociated by a
continuous-wave UV laser (266 nm) using different intensities. The
molecules were sympathetically cooled by simultaneously trapped laser-cooled
barium ions to translational temperatures of below 150 mK. Destruction
rates of less than 0.05 s$^{-1}$ and a cross section of $(1.1\pm0.1)\cdot10^{-17}$
cm$^{2}$ have been determined. An extension to tunable UV laser sources
would permit high-resolution dissociation spectroscopic studies on
a wide variety of cold complex molecules.

\maketitle
\end{abstract}

\section{Introduction}

Photodissociation spectroscopy is a common tool to gain information
on structures of biomolecules \cite{Correia2007}, to distinguish
between different isomers \cite{Stearns2007} or to study the energetics
and pathways of fragmentations \cite{Dienes1996}. The investigation
of these processes in the gas phase provides advantageous conditions.
Here, one can guarantee, by mass selection prior to spectroscopy,
the sample does not contain dimers, aggregates or chemically modified
molecules. In the low-pressure environment of the gas phase, the collision
rate can be strongly reduced. This enables the study of the above
processes with little influence from intermolecular interactions.
The interpretation of the results by quantum chemical calculations
will therefore generally be facilitated \cite{Nolting2004}. 

Trapped-ion photodissociation offers additional advantages. The long
storage times in ion traps can extend the time scale of observable
dissociation processes, such as blackbody infrared radiative dissociation
(BIRD) \cite{Price1996} or the unimolecular dissociation of large
biomolecules at low energies \cite{Griffin1993,Schlag1989}, to rates
smaller than 1 s$^{-1}$ which cannot be obtained using other, non-trapping
approaches \cite{Dunbar2000,Worm2007,Stoechkel2008}. Additionally,
due to the strong spatial localization of the trapped ions, even single
molecules can be studied \cite{Hoejbjerre2008} resulting in yet simplified
environmental conditions. By cooling the trapped molecules, spectral
congestion and inhomogeneous line broadening can be reduced and photodissociation
spectra of high resolution can be obtained. A conventional way of
cooling both internal and external degrees of freedom of molecular
ions in traps is using cryogenic buffer gases \cite{Stearns2007a,Koelemeij2007}
leading to translational and internal temperatures of a few Kelvin,
but no strong spatial confinement. A combination of sympathetic translational
cooling via laser cooled atomic ions \cite{Ostendorf2006,Offenberg2008}
and internal cooling might be feasible via radiative cooling in a
cryogenically cooled ion trap \cite{Berkeland1997}. This promises
to reach both much lower temperatures than with the conventional buffer
gas method and simultaneously a stronger spatial confinement.

Few quantitative determinations of photodestruction rates have been
performed. One of the earliest was the measurement of the wavelength-dependent
photodestruction of CH$_{4}^{+}$ in a Paul trap, where cross sections
in the order of $1\cdot10^{-19}$ cm$^{2}$ were determined \cite{Ensberg1975,Khoury2002}.
Here, we demonstrate a quantitative measurement of the photodestruction
rate of a large polyatomic molecular ion species. The measurement
of the cross section was performed as a function of laser intensity,
allowing to verify that only one-photon processes contribute. 

We employ an apparatus that provides protonated molecular ions with
masses up to 12400 Da produced by electrospray ionization (ESI) at
sub-Kelvin translational temperatures \cite{Ostendorf2006,Offenberg2008}.
For this purpose, the molecular ions are stored in a linear quadrupole
trap and are sympathetically cooled by simultaneously trapped laser-cooled
$^{138}$Ba$^{+}$ ions via their mutual Coulomb interaction. The
organic compound glycyrrhetinic acid (mass 471 Da) is used as a test
molecule as it is chemically stable, inexpensive and available at
a reliably high ESI ion flux. A low-power, fixed-frequency continuous-wave
(cw) UV laser at 266 nm was used to demonstrate two different methods
for determining photodestruction rates of cold molecules. The first
method is based on a continuous monitoring of the decay of the parent
ion number during the dissociation. In the second method, the destruction
process is characterized by extracting and counting the ions remaining
after different UV exposure durations.

\begin{figure}[b]
\begin{centering}
\includegraphics{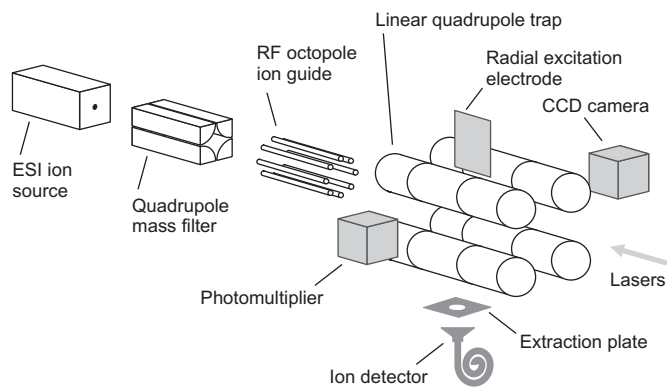}
\par\end{centering}

\caption{Schematic overview of the experimental setup. Protonated molecular
ions from an ESI ion source are selected by a quadrupole mass filter
and transferred via a RF octopole ion guide to a linear quadrupole
trap in a UHV chamber. The trapped ions can be electronically excited
by a radially mounted electrode and counted by an ion detector below
the trap. The laser-cooled $^{138}$Ba$^{+}$ ions are imaged with
a CCD camera and their fluorescence is detected by a photomultiplier.}
\label{fig:1}
\end{figure}

\section{Experimental setup and basic principles}

Our experimental setup \cite{Ostendorf2006,Offenberg2008} is shown
schematically in figure \ref{fig:1}. It consists of an ESI ion source
\cite{Fenn1989} for the production of singly or multiply charged,
gas-phase molecular ions with a mass-to-charge ratio $m/z$ ($z$
is the number of elementary charges) of up to 2000 Da, a quadrupole
mass filter for the selection of specific molecular species, a RF
octopole ion guide to transfer the selected molecular ions from the
medium vacuum region of the ESI device to an ultra-high vacuum (UHV)
chamber with a typical residual gas pressure of $<10^{-9}$ mbar,
and a linear quadrupole trap in this UHV chamber to store the molecular
ions for sympathetic cooling and the intended experiments. 

\begin{figure}
\begin{centering}
\includegraphics{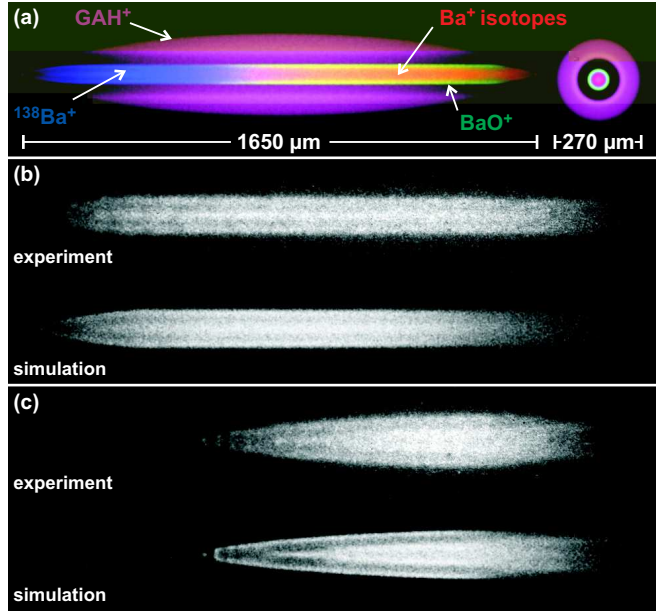}
\par\end{centering}

\caption{Experimental and simulated CCD images of a Ba$^{+}$/GAH$^{+}$ ion
crystal. (a) Simulated images in radial (left) and axial view (right),
as they would appear if all ions would fluoresce. The cooling lasers
propagate to the left and separate the $^{138}$Ba$^{+}$ ions (blue)
and the barium isotopes (red) due to the light pressure force. The
GAH$^{+}$ ions (pink) form a sheath around the barium ion subensemble.
The BaO$^{+}$ ions (green) are impurities which in this case were
not completely removed. (b) Experimental image of the Ba$^{+}$/GAH$^{+}$
ion crystal and its simulation, showing only the fluorescing $^{138}$Ba$^{+}$
ions. (c) Crystal after removal of the GAH$^{+}$ ions. Here, the
barium ion subensemble is no longer deformed nor heated by the GAH$^{+}$
ions.}
\label{fig:2}
\end{figure}

The preparation of the laser-cooled $^{138}$Ba$^{+}$ ion ensembles
used for the sympathetic cooling of the molecular ions as well as
the required laser setup (a 493 nm cooling and a 650 nm repumper laser)
have been described in \cite{Roth2005}. Under appropriate laser-cooling
conditions, the barium ions arrange in ordered structures, so-called
Coulomb crystals, that can be imaged with an intensified CCD camera
(see figures \ref{fig:2} (b) and (c)). The images contain information
on numbers and temperatures of both the fluorescing $^{138}$Ba$^{+}$
ions and indirectly of the invisible sympathetically cooled molecular
ions. Using molecular dynamics (MD) simulations, these data can be
derived from structural deformations and blurring of the barium ion
Coulomb crystals \cite{Zhang2007}. Further components for the detection
of the trapped ions are an ion detector to count mass-selectively
extracted ions and a radially mounted electrode for the excitation
of species-specific oscillation modes. A resonant electric excitation
causes a drop of the barium ions' fluorescence that can be measured
with a photomultiplier. 

Ensembles of cold ($<1$ K), singly protonated glycyrrhetinic acid
molecules (GAH$^{+}$) produced by ESI are routinely prepared within
less than 1 min by the following procedure \cite{Ostendorf2006,Offenberg2008}.
Here, with a glycyrrhetinic acid (Sigma-Aldrich) solution of $10^{-5}$
M concentration in 1:1 acetonitrile:water with $5\cdot10^{-3}\%$
formic acid added for protonation, a flux of several ten thousand
GAH$^{+}$ ions per second leaving the octopole ion guide is achieved.
This flux can be switched on and off via the octopole RF voltage.
For trapping the molecular ions, helium buffer gas at room temperature
is injected into the trap chamber to reduce the ions' kinetic energy.
The number of trapped ions depends linearly on the buffer gas pressure
and the duration of the ion flux. For typical pressures of $1-10\cdot10^{-5}$
mbar several hundred molecular ions are trapped within a few seconds.
After removal of the buffer gas, molecular fragment ions generated
during loading are removed from the trap by applying an additional
AC voltage frequency scan to the trap electrodes in order to excite
the unwanted ions' specific motional resonances so strongly that these
ions are ejected from the trap. Then, barium ions from an evaporator
oven are loaded into the trap and cooled by the two cooling laser
beams propagating along the trap axis. After two more frequency scans
that remove impurity ions (such as CO$_{2}^{+}$ and BaO$^{+}$) generated
during loading the barium, the preparation of a cold Ba$^{+}$/GAH$^{+}$
ion crystal is completed.

Figure \ref{fig:2} shows experimental and simulated CCD images of
such an ion crystal revealing the spatial configuration of the different
ion species. In a quadrupole trap, ions of different charge-to-mass
ratios $Q/m$ arrange radially separated according to their effective
radial trap potentials \begin{equation}
\Phi(r)\propto\frac{Q}{m}r^{2}\label{eq:radial potential}\end{equation}
and the interspecies repulsion $\sim Q_{1}Q_{2}$ \cite{Offenberg2008}.
For prolate crystals, this radial separation scales as \begin{equation}
\frac{r_{1}}{r_{2}}\simeq\sqrt{\frac{m_{1}/Q_{1}}{m_{2}/Q_{2}}}\label{eq:radial separation}\end{equation}
with the outer radius $r_{1}$ of the lower mass-to-charge ratio $m_{1}/Q_{1}$
subensemble and the inner radius $r_{2}$ of the higher mass-to-charge
ratio $m_{2}/Q_{2}$ subensemble \cite{Wineland1987}. Thus, the heavier
GAH$^{+}$ ions arrange around the barium ions and form a sheath which
radially squeezes and axially prolongs the barium ion subensemble
as shown in figure \ref{fig:2}. From this structural deformation
the number of trapped GAH$^{+}$ ions can be derived. Therefore, two
experimental images are recorded $-$ one with the molecular ions
present and one after the molecular ions have been removed. In the
case shown in figure \ref{fig:2}, the GAH$^{+}$ ions prolong the
barium ion subensemble by 252 \textmu{}m and radially squeeze it by
25 \textmu{}m. According to MD simulations this can be explained by
the presence of $1250\pm50$ GAH$^{+}$ ions. The barium ion subensemble
itself consists of $240\pm10$ laser-cooled $^{138}$Ba$^{+}$ ions
and $165\pm10$ barium isotopes%
\footnote{Natural barium consists of seven stable isotopes, the most abundant
is $^{138}$Ba with a proportion of $72\%$. In our simulations the
other six isotopes are treated as a single species with a weighted
mass of 136 amu. However, in experimentally observed ion crystals
the isotope proportion can be significantly larger than $28\%$ due
to a loss of electronically excited $^{138}$Ba$^{+}$ ions by laser
assisted chemical reactions with residual gas O$_{2}$ and CO$_{2}$
molecules that are energetically not possible with the ground state
isotope ions \cite{Roth2008}.%
} that arrange on the right side of the crystal due to the light pressure
force of the cooling lasers propagating to the left and only acting
on the laser-cooled species. An additional structural detail in this
discussed case is the slight radial constriction on the right side
of the barium ion crystal (see figure \ref{fig:2} (c)). This can
be ascribed to BaO$^{+}$ or BaOH$^{+}$ ions (in this case $150\pm10$
ions) that are generated by photo-induced reactions between $^{138}$Ba$^{+}$
ions in the 6$^{2}$P$_{1/2}$ excited state and neutral CO$_{2}$,
O$_{2}$ or H$_{2}$O molecules from residual gas \cite{Roth2008},
which was present at an unintendedly high partial pressure in this
case. However, under normal operating vacuum conditions ($p<10^{-9}$
mbar) the reaction rates are low and the loss of $^{138}$Ba$^{+}$
is negligible.

From the sharpness and structural details of the barium ion crystals
the translational temperatures $T_{sec,i}=\frac{2}{3}\left\langle E_{i}\right\rangle /k_{B}$
of the ions can be deduced using MD simulations, where $\left\langle E_{i}\right\rangle $
is the time- and subensemble-averaged secular kinetic energy per ion
of the species $i$ \cite{Zhang2007}. For the fluorescing $^{138}$Ba$^{+}$
ions the temperatures can be obtained by a direct comparison of experimental
and simulated CCD images, whereas for the sympathetically cooled invisible
ions the temperatures can be derived indirectly via the temperature
increase of the barium ions due to the sympathetic interaction. As
the laser cooling efficiency varies from case to case depending on
numerous factors, we consider a realistic range of laser-cooling rates
in the simulations in order to obtain the most probable translational
temperatures together with a lower and an upper limit. In the discussed
case, the GAH$^{+}$ ions had a temperature of $134{}_{-24}^{+8}$
mK, the Ba$^{+}$ ions $74$ mK, the barium isotope ions $122{}_{-20}^{+5}$
mK and the BaO$^{+}$ ions $119{}_{-19}^{+5}$ mK. We expect that
GAH$^{+}$ temperatures below 100 mK could be achieved by reducing
the ratio of the numbers of GAH$^{+}$ and Ba$^{+}$ ions and the
total number of ions, operating at lower residual gas pressures or
under improved laser-cooling conditions.

\section{Photodestruction of GAH$^{+}$}

\begin{figure}
\begin{centering}
\includegraphics{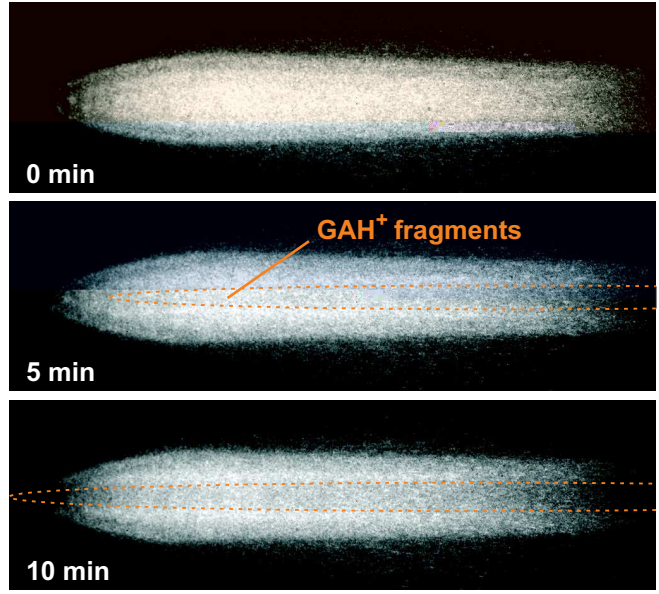}
\par\end{centering}

\caption{Observation of GAH$^{+}$ photodestruction. An ion crystal consisting
of laser-cooled $^{138}$Ba$^{+}$ ions (visible in the CCD image),
barium isotopes, BaO$^{+}$ and GAH$^{+}$ (all three not visible)
is continuously exposed to 266 nm laser light of low intensity. The
GAH$^{+}$ ions dissociate within minutes, their fragments accumulate
along the trap axis forming a dark core (dashed lines) inside the
barium ion subensemble.}
\label{fig:3}
\end{figure}

In the experiment we observe that a Ba$^{+}$/GAH$^{+}$ ion crystal
exposed to 266 nm cw laser light changes its shape due to the destruction
of the GAH$^{+}$ ions as shown in figure \ref{fig:3}. The produced
fragments have lower mass-to-charge ratios than the barium ions and
therefore accumulate closer to the trap axis forming a dark core in
the barium ion subensemble, which grows during the dissociation process.
The fragments' mass-to-charge ratios range between 40 and 82 Da as
determined by a mass-selective extraction of the ions from the trap
(see section \ref{sub:Extraction-method}). A destruction of GAH$^{+}$
by the barium cooling lasers (493 and 650 nm) alone has not been observed.
Here, we show two methods to measure quantitatively the decay of the
GAH$^{+}$ parent ion number. In the \char`\"{}excitation method\char`\"{}
the decay is detected during one photodestruction process via a repeated
excitation of the parent ions that produces signals proportional to
the number of parent ions. In the \char`\"{}extraction method\char`\"{}
the composition of a set of initially similar Ba$^{+}$/GAH$^{+}$
ion crystals is analyzed via extractions of the ions from the trap
after different times of UV exposition. The analysis of the GAH$^{+}$
parent ion number decay relies on the rate equation model derived
in section \ref{sub:Discussion}. In this model, the parent ion number
$N(t)$ after a UV exposure time $t$ is given by\begin{equation}
N(t)=N(0)\exp\left(-\gamma t\right)\label{eq:number(t)}\end{equation}

with the photodestruction rate $\gamma$. Note that this rate is not
a photodissociation rate in the sense of an inverse lifetime of an
excited molecule against dissociation. Rather, it is determined by
the absorption rate times the sum of fractional probabilities for
various dissociation pathways, such as unimolecular, statistical,
and nonstatistical dissociation. The various pathways cannot be specified
further due to the principle of our measurements.

\subsection{Excitation method}

In the harmonic, effective electric potential of a linear quadrupole
trap \cite{Broener2000}, ions can oscillate radially with a mass-to-charge
ratio specific resonance frequency\begin{equation}
\omega_{r}=\frac{\Omega}{2}\sqrt{\frac{q^{2}}{2}+a}\label{eq:radial frequency}\end{equation}

with the Mathieu stability parameters\begin{equation}
q=\frac{2QU_{\mathrm{RF}}}{mr_{0}^{2}\Omega^{2}}\label{eq:q-parameter}\end{equation}

and

\begin{equation}
a=\frac{-4\kappa QU_{\mathrm{EC}}}{m\Omega^{2}}.\label{eq:a-parameter}\end{equation}

Typical trap parameters in our case are $U_{\mathrm{RF}}=200-500$
V for the trap RF amplitude, $U_{\mathrm{EC}}=5-7$ V for the DC potential
difference between the end and middle segments of the trap electrode
rods (see figure \ref{fig:1}). Constant parameters are the RF frequency
$\Omega=2\pi\cdot2.5$ MHz, the axis-to-electrode-surface distance
$r_{0}=4.36$ mm, and the geometrical factor $\kappa=1500$ m$^{-2}$. 

These oscillations can be excited by AC voltages applied to the trap
electrodes or an additional excitation electrode (see figure \ref{fig:1}).
When the excitation frequency coincides with the resonance frequency
of an ion species, its strong large-amplitude response heats up all
trapped ions due to their mutual Coulomb interaction. As a consequence,
the barium ions' fluorescence drops, leading to dips in the recorded
fluorescence signal as a function of the AC frequency \cite{Baba2002,Roth2007}.
The depth of theses dips is approximately proportional to the number
of excited ions \cite{Roth2006}. This was confirmed by comparison
of the results of the two methods presented here (see section \ref{sub:Discussion}). 

\begin{figure}
\begin{centering}
\includegraphics{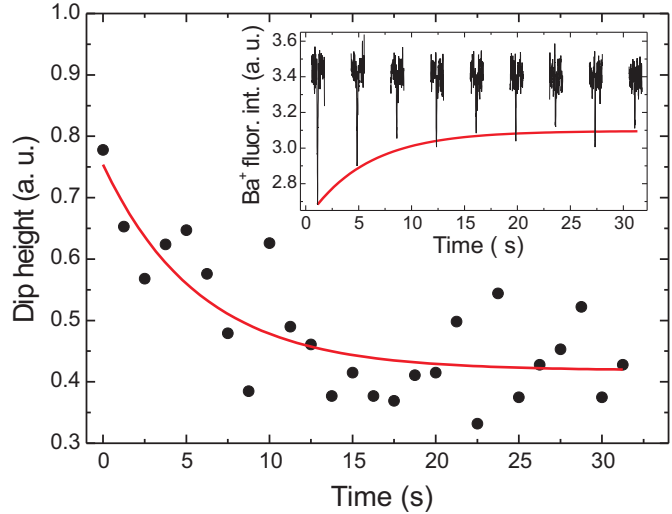}
\par\end{centering}

\caption{Photodestruction of GAH$^{+}$ measured with the excitation method.
The inset shows repeated radial excitation scans over the GAH$^{+}$
resonance with their height decreasing during the photodissociation
of the GAH$^{+}$ parent ions by a 266 nm laser at 12 mW/cm$^{2}$.
Due to the occurrence of a new dip in the scanned frequency range,
the dips do not vanish completely. The main plot shows the dip heights
of the complete measurement with an exponential fit. The average of
six such measurements at this intensity yields a rate of $\gamma=(0.15\pm0.02)$
s$^{-1}$.}
\label{fig:4}
\end{figure}

Thus, the decay of the number of a sympathetically cooled ion species
can be followed by recording the barium ions' fluorescence during
repeated frequency scans over the species' resonance. The result of
such a measurement is shown in figure \ref{fig:4} for the photodestruction
of GAH$^{+}$ with 266 nm laser light at an intensity of 12 mW/cm$^{2}$.
The radial resonance of GAH$^{+}$ was repetitively excited via the
external excitation electrode by frequency scans from 35 to 15 kHz
with a duration of 0.2 s and a repetition rate of 0.8 Hz. For clarity,
the inset of figure \ref{fig:4} shows only every third resonance
dip of this destruction process, while the dip heights of the complete
measurement are shown in the main plot. An exponential fit to the
decay of the GAH$^{+}$ resonance dip height, which equals the decay
of the GAH$^{+}$ parent ion number, yields the destruction rate $\gamma$.
The average of six such measurements at this intensity of 12 mW/cm$^{2}$
yields a rate of $\gamma=(0.15\pm0.02)$ s$^{-1}$. Several measurements
with different dissociation laser intensities have been performed
both with this and the extraction method; the results are discussed
later and shown in figure \ref{fig:7}.

Unfortunately, for this test molecule GAH$^{+}$ and probably for
other molecules with similar masses ($471\pm50$ Da) this method is
ambiguous. With the parent ion number decreasing to zero, their resonance
dips should vanish completely during the dissociation process. However,
we cannot observe this for GAH$^{+}$ (see figure \ref{fig:4}). Here,
during the dissociation a new resonance arises close to the parent
ion resonance which cannot be resolved and just slightly broadens
the observed parent ion dip. We can explain this new resonance as
being due to the generated fragment ions. Their mass-to-charge ratios
of 40 to 82 Da lead not only to radial resonances between 260 to 130
kHz as expected from equation \ref{eq:radial frequency} with $U_{\mathrm{RF}}=291$
V as applied in the case of figure \ref{fig:4}, but also to an additional
resonance which coincidentally agrees with that of the parent ions
($\approx$ 22 kHz in this case). This can be explained by a motional
coupling of all trapped ions \cite{Roth2007}. Even so, an evaluation
of the rates of the GAH$^{+}$ dissociation process measured with
this method was possible (as in figure \ref{fig:4}) and lead to results
that agree well with those measured with the extraction method (see
figure \ref{fig:7}, open circles). For other parent ion masses sufficiently
different from that of GAH$^{+}$, this problem will not occur, as
their resonances can be resolved from those arising from their fragments.
Generally speaking, the mass resolution of radial excitation depends
on the numbers and mass-to-charge ratios of the trapped ion species
of a specific ion ensemble and is limited by motional coupling effects.
The mass resolution is in the order of about 10\% but coupling effects
can lead to ambiguous results as mentioned above.

\subsection{Extraction method \label{sub:Extraction-method}}

\begin{figure}[t]
\begin{centering}
\includegraphics{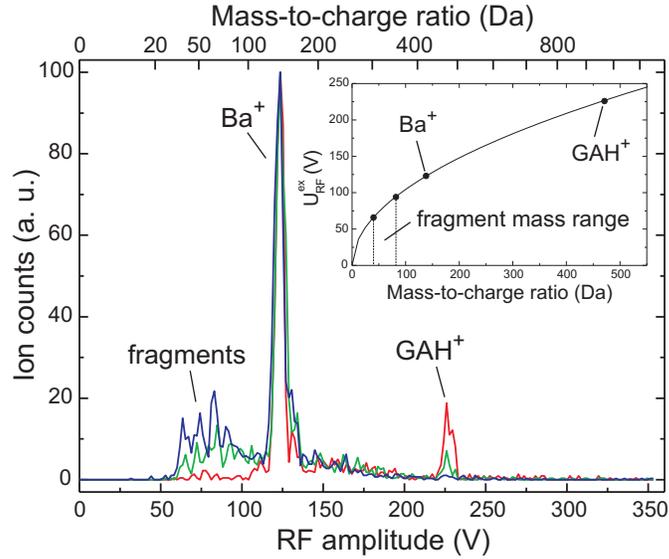}
\par\end{centering}

\caption{Ion extraction mass spectra of Ba$^{+}$/GAH$^{+}$ crystals at different
stages of GAH$^{+}$ dissociation. The red curve is the extraction
spectrum of an intact Ba$^{+}$/GAH$^{+}$ crystal showing only the
two species. The green curve shows the extraction of an identically
prepared crystal after a 1.2 s exposition to 266 nm laser light at
55 mW/cm$^{2}$. Here, the number of GAH$^{+}$ ions is reduced and
fragments appear. After 2.4 s (blue curve) all GAH$^{+}$ ions are
dissociated and the number of fragments reaches a final maximum. The
inset curve shows the theoretical dependency of the extraction RF
amplitude $U_{\mathrm{RF}}^{\mathrm{ex}}$ on the mass-to-charge ratio,
calibrated with the known mass of barium. The GAH$^{+}$ data point
indicates that indeed intact GAH$^{+}$ molecules have been stored.}
\label{fig:5}
\end{figure}

In the extraction method we make use of a mass-to-charge ratio selective
extraction of the ions from the trap. When decreasing the trap RF
amplitude $U_{\mathrm{RF}}$ the Mathieu stability parameter $q$
(see equation \ref{eq:q-parameter}) is reduced and the ions escape
from the trap at a mass-to-charge ratio dependent amplitude \begin{equation}
U_{\mathrm{RF}}^{\mathrm{ex}}=\alpha\cdot r_{0}^{2}\Omega\sqrt{2\kappa U_{\mathrm{EC}}\frac{m}{Q}}.\label{eq:extractionamp}\end{equation}
This expression with $\alpha=$1 follows from the Mathieu stability
diagram (e. g. shown in \cite{Broener2000}) with regard to the decrease
of $q$ below stability%
\footnote{This holds when the edge of the stability range is approximated by
$q=-\frac{1}{2}a^{2}$ which is adequate for the range of $q$ considered
here.%
}. In our setup, an extraction plate (see figure \ref{fig:1}) 17 mm
below the trap center at a potential of $-1000$ V referred to ground
draws the escaped ions to an ion detector that registers these ions
during such a controlled reduction of the RF amplitude, leading to
an ion extraction mass spectrum as shown in figure \ref{fig:5}. The
overall ion capture efficiency is between 10 and 20 \% depending mainly
on the applied detector electronics settings. During the experiments
the efficiency is constant and almost equal for the various singly
charged species. This was determined by comparing ion numbers detected
by the counter with the numbers in the Coulomb crystals determined
by molecular dynamics simulations \cite{Zhang2007}. The numerical
factor $\alpha>1$ arises from the attracting effect of the extraction
plate that causes the ions to escape the trap earlier during the RF
amplitude reduction, i. e. at higher amplitudes. Its value slightly
varies depending on ion numbers and temperatures and is approximately
2.4 in our setup. Using the barium peak in an ion extraction mass
spectrum for a precise calibration of $\alpha$, we find that we can
then verify the mass of GAH$^{+}$ and determine the mass-to-charge
ratio range of its fragments to 40 $-$ 82 Da (see figure \ref{fig:5}).

This mass identification technique is destructive, but it has several
advantages compared to the non-destructive radial excitation. It unambiguously
distinguishes species according to their mass-to-charge ratio reaching
a resolution $<20$ Da below $\sim500$ Da for cold ensembles. Furthermore,
it is also applicable to uncooled ensembles, however with a lower
resolution, and it is fast because an extraction only takes a few
seconds without any preparation. 

In order to measure photodestruction rates with this technique, ion
crystals prepared under identical conditions containing the desired
molecular ions are exposed to the dissociation laser for different,
sequentially increased times. The resulting ion crystals in different
stages of the destruction process are analyzed via ion extraction.
Thus, the decay of the parent ion number is fully described by such
a set of ion extraction spectra. Figure \ref{fig:5} shows ion extraction
spectra of identically prepared Ba$^{+}$/GAH$^{+}$ crystals in different
stages of the GAH$^{+}$ dissociation by 266 nm laser light of 55
mW/cm$^{2}$ intensity. The red curve is the extraction spectrum of
an intact Ba$^{+}$/GAH$^{+}$ crystal acquired directly after preparation
showing only the two species. The green curve was acquired after an
UV exposition of 1.2 s and shows a reduced number of GAH$^{+}$ ions
and the appearance of fragments. After 2.4 s all GAH$^{+}$ ions are
dissociated and the number of fragments reaches a final maximum as
shown in the blue curve.

\begin{figure}[b]
\begin{centering}
\includegraphics{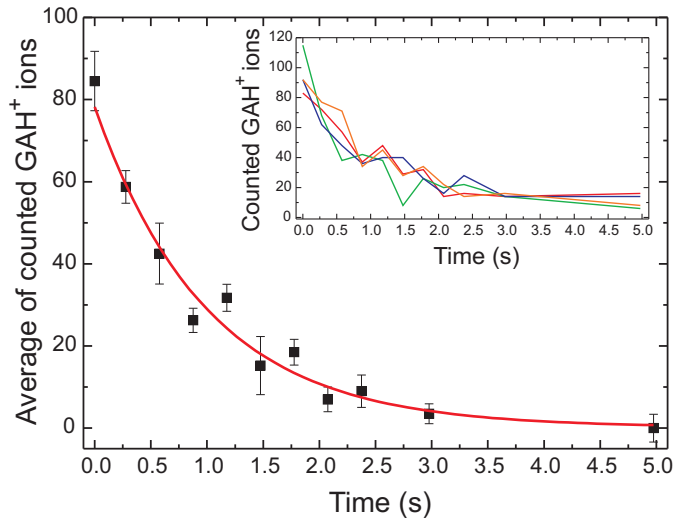}
\par\end{centering}

\caption{Photodissociation of GAH$^{+}$ measured with the extraction method
for an exposition with 266 nm laser light at 55 mW/cm$^{2}$. The
inset shows the decay of the GAH$^{+}$ parent ion number of four
sets of 11 ensembles, where the crystals of each set were UV-exposed
consecutively for increasing durations. The main plot shows the decay
of the GAH$^{+}$ numbers, averaged over the four sets shown in the
inset. The exponential fit yields a photodestruction rate $\gamma=(1.00\pm0.13)$
s$^{-1}$.}
\label{fig:6}
\end{figure}

A parent ion decay curve is obtained from such a set of ion extraction
spectra by adding up the ion counts of the parent ion peak in each
extraction spectrum. However, although all starting ion crystals of
a dissociation run are prepared under identical conditions, there
are deviations in the numbers of loaded molecular ions with a standard
deviation of up to $28\%$. This is due to the irregular molecular
ion flux of the ESI ion source that shows as similar standard deviation
of $25\%$, a typical value for ESI sources \cite{RizzoPC2008}. Therefore,
several complete dissociation runs are measured for the same spectroscopy
condition (laser intensities, wavelengths) in order to minimize this
non-systematic error. In the discussed case (see above and figure
\ref{fig:5}) four GAH$^{+}$ dissociation runs have been acquired
leading to the decay curves shown in the inset of figure \ref{fig:6}.
The main plot shows the averaged data points of these curves and an
exponential fit that yields a rate $\gamma=(1.00\pm0.13)$ s$^{-1}$
for this photodestruction process.

\subsection{Discussion\label{sub:Discussion}}

The two presented methods are based on fundamentally different principles
and therefore show diverse advantages and difficulties. The most striking
difference is the time it takes to measure a destruction rate for
a given spectroscopy condition. Under best experimental conditions,
such a measurement takes a few minutes with the excitation method,
whereas it takes at least one hour with the extraction method. This
is because in the latter case dozens of ion crystals need to be prepared
and destructed to achieve lower relative uncertainties, compared to
only one crystal in the first case. 

However, this speed difference is often compensated by the higher
experimental effort of the excitation method. Here, it is a complex
task to produce ion crystals with a certain required critical ratio
of numbers of laser-cooled atomic ions and sympathetically cooled
molecular ions. Two conditions need to be balanced $-$ the signal-to-noise
ratio of the resonance dips needs to be sufficiently high and the
ion ensembles need to stay in the crystal state during the whole dissociation
process. The first condition, a high signal-to-noise ratio, can be
easily achieved with a higher number of molecular ions that is only
limited by the finite sympathetic cooling power of the laser-cooled
ions. However, the maintenance of the crystal state is more critical,
because every resonant excitation of the ions implies a disturbance.
When the number of molecular ions is too large, the ion crystals melt
before the dissociation process is finished, as they are perturbed
by the produced fragment ions due to their stronger sympathetic interaction
with the laser-cooled ions. In addition, with higher numbers of molecular
ions and their fragments, the ion crystals are more sensitive, so
that slightly stronger or more frequent excitations can easily melt
the crystals. 

Thus, the preparation and the handling of barium/molecule ion crystals
for the application of the excitation technique requires a precise
and critical control of the loading and excitation parameters. In
comparison, the extraction method is less sensitive but more straightforward,
as it does without the crucial excitation. Here, the only condition
to be fulfilled, is to prepare ion crystals with a sufficient number
of molecular ions that produces a distinct peak in the ion extraction
spectra. We expect that this method could be automated to a large
extent for the application in spectroscopic studies.

\begin{figure}
\begin{centering}
\includegraphics{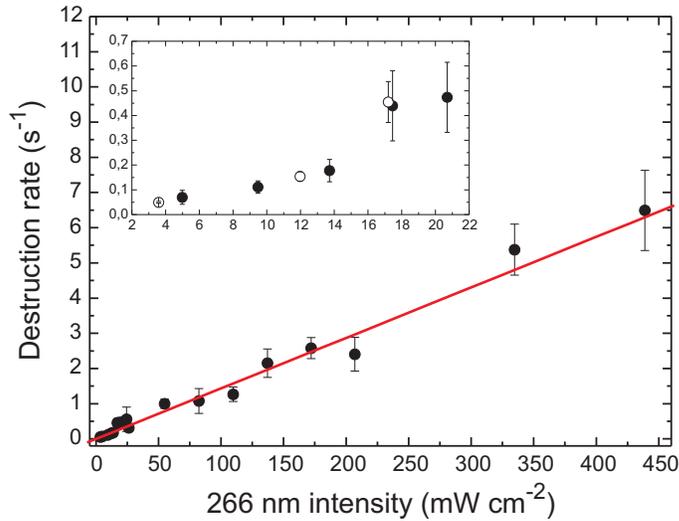}
\par\end{centering}

\caption{Destruction of translationally cold GAH$^{+}$ at different laser
intensities. The destruction rates shows a linear dependence on the
266 nm dissociation laser intensity in the range of intensities covered
(the red line is a linear fit). The rates measured with the excitation
method (open circles, inset) agree well with those measured with the
extraction method (filled circles).}
\label{fig:7}
\end{figure}

To demonstrate the practicability of our methods, we have investigated
the photodestruction of GAH$^{+}$ with a cw 266 nm laser at low intensities
ranging from 3 to 440 mW/cm$^{2}$. Figure \ref{fig:7} shows the
photodestruction rates obtained with the extraction method (filled
circles) and the excitation method (open circles, inset). The measured
rates range from 0.05 to 6.5 s$^{-1}$ and linearly depend on the
dissociation laser intensity. Significantly lower rates should in
principle be observable due to the long-term stable storage of the
molecular ions of up to hours. Measurements of rates higher than $\sim$
0.5 s$^{-1}$ are not possible with the excitation method as the excitation
repetition rate is limited to $\lesssim$ 2 Hz. With higher repetition
rates the crystals cannot cool down sufficiently after excitation
to maintain crystallization. In contrast, for the extraction method
the upper limit of measurable rates is only limited by the shutting
time of the dissociation lasers which is in the range of milliseconds
using mechanical shutters or nanoseconds for electro-optical switches. 

With our apparatus we cannot and do not intend to give insight into
the precise nature of the destruction process of GAH$^{+}$ or other
molecular species. Our mass resolution is too low to distinguish all
generated fragment ion species. In addition, the further dissociation
of fragments due to the comparatively long UV exposure times in the
measurements presented here complicates any kind of interpretation.
Various pathways for the unimolecular dissociation are possible like
loss of neutral or charged groups or cleavage of the molecules at
weak bonds \cite{Worm2007}. Generally, gas-phase biomolecules of
sizes comparable to that of GAH$^{+}$ can dissociate after the absorption
of a single UV photon of 266 nm \cite{Worm2007,Liu2006,Nielsen2004},
which we well confirmed by the linear intensity dependence of the
measured destruction rates of GAH$^{+}$. After absorption, such molecules
either dissociate within less than microseconds from an electronically
excited state (nonstatistical dissociation) or on a timescale of milliseconds
after redistribution of the internal energy to vibrational modes (statistical
dissociation). Especially for protonated molecules of sizes comparable
to that of GAH$^{+}$, the \char`\"{}immediate\char`\"{} nonstatistical
dissociation is the dominant process \cite{Nielsen2004}. Based on
this assumption the destruction process can be described by a simple
rate equation model as motivated in \cite{Khoury2002} given by

\begin{equation}
\frac{dN_{e}(t)}{dt}=N_{g}(t)\cdot R-N_{e}(t)\cdot\left(R+\Gamma+k\right)\label{eq:ne(t)}\end{equation}

\begin{equation}
\frac{dN_{g}(t)}{dt}=-N_{g}(t)\cdot R+N_{e}(t)\cdot\left(R+\Gamma\right)\label{eq:ng(t)}\end{equation}
Here, GAH$^{+}$ ions in the electronic ground state with an ion number
$N_{g}(t)$ absorb laser photons with the absorption rate $R=\sigma_{abs}I/h\nu$,
where $\sigma_{abs}$ is the absorption cross section of GAH$^{+}$
and $I$ the intensity of the dissociation laser at the frequency
$\nu=c/(266$ nm). $N_{e}(t)$ is the ion number in the electronically
excited state. Its decay is characterized by the fluorescence rate
$\Gamma$ and its photodissociation channel by the unimolecular dissociation
rate $k$. The solution for the total parent ion number $N(t)=N_{g}(t)+N_{e}(t)$
in the relevant case $R\ll k,\Gamma$ at the low intensities $I$
used in our measurements is given by\begin{equation}
N(t)=N(0)\cdot\exp\left(-\gamma t\right)\label{eq:n(t)}\end{equation}

with the photodestruction rate\begin{equation}
\gamma=\frac{kR}{k+\Gamma}.\label{eq:gamma}\end{equation}

With the relation $\gamma=\sigma\cdot I/h\nu$ valid for single-photon
processes, we define a photodestruction cross section\begin{equation}
\sigma_{pd}=\sigma_{abs}\frac{k}{k+\Gamma}.\label{eq:crosssection}\end{equation}

A linear fit to the data of our measurement shown in figure \ref{fig:7}
yields a photodestruction cross section at 266 nm of translationally
cold, trapped GAH$^{+}$ ions of $\sigma_{pd}=(1.1\pm0.1)\cdot10^{-17}\mathrm{cm^{2}}$. 

Only few quantitative measurements on UV photodestruction cross sections
of trapped molecular ions have been performed. Similar measurements
on trapped, but warm CH$_{4}^{+}$ \cite{Ensberg1975} yielded smaller
photodestruction cross sections of the order of $1\cdot10^{-19}$
cm$^{2}$ for visible wavelengths, with a tendency to increase with
decreasing wavelength. In our model, for high photodissociation rates
$k\gg\Gamma$, the photodestruction cross section $\sigma_{pd}$ would
equal the absorption cross section $\sigma_{abs}$, while in any other
case it would be less. There are, to our knowledge, no quantitative
measurements on the UV absorbtion cross section of GAH$^{+}$ so that
we can only compare the lower limit $(\sigma_{abs})_{min}=\sigma_{pd}$
with absorption cross sections of other gas-phase organic molecules.
For example, a gas-phase amide (C$_{4}$H$_{9}$NO) shows an absorption
cross section at 266 nm of $\sigma_{abs}(266\mathrm{nm})\approx5\cdot10^{-19}$
cm$^{2}$ \cite{Chakir2005} and neopentylperoxy radical molecules
(C$_{5}$H$_{11}$O$_{2}$) $\sigma_{abs}(266\mathrm{nm})\approx4\cdot10^{-18}$
cm$^{2}$ \cite{Dagaut1990}, that are both lower than that of GAH$^{+}$.
Measurements on fullerenes (C$_{60}$) at the same wavelength gave
$\sigma_{abs}(266\mathrm{nm})\approx4\cdot10^{-16}$ cm$^{2}$ \cite{Smith1996}.

\section{Conclusion}

In this work, we have shown two methods for the measurement of photodestruction
rates of translationally cold, trapped molecular ions. A particular
advantage of our technique are the long storage times of many minutes
and in principle of up to hours in the well-defined and nearly collisionless
environment of an ion trap in an ultrahigh vacuum. This allows for
the study of slow destruction processes such as the photodissociation
of large biomolecules. For trapped and translationally cooled, singly
protonated ions of the organic compound glycyrrhetinic acid a photodestruction
cross section at 266 nm of $(1.1\pm0.1)\cdot10^{-17}\mathrm{cm^{2}}$
has been determined. In future, the methods could be applied for a
systematic acquisition of highly resolved photodissociation spectra
using low intensity, tunable cw lasers with narrow linewidths. In
order to reduce the spectral congestion due to the presence of numerous
conformers, a cooling of the internal degrees of freedom would be
highly advantageous and could be implemented by radiative cooling
in a cryogenic environment \cite{Berkeland1997}.

\section*{Acknowledgments}

We thank R. Edrada-Ebel and R. Weinkauf for helpful discussions. DO
acknowledges support from the Studienstiftung des deutschen Volkes
and CBZ from the Deutscher Akademischer Austauschdienst (DAAD). This
work was supported by the German Science Foundation.

\end{document}